\documentclass[aps,showpacs,preprint,preprintnumbers,nofootinbib,tighten]{revtex4}
\usepackage{graphicx}
\usepackage{dcolumn}
\usepackage{bm}
\usepackage{amssymb,amsmath}
\usepackage{epsfig}                                                             

\begin{document}

\title{\Large{Ratio of viscosity to entropy density
in a strongly coupled one-component plasma}}

\author{Markus H. Thoma and Gregor E. Morfill}

\affiliation{Max-Planck-Institut f\"ur extraterrestrische Physik,
P.O. Box 1312, 85741 Garching, Germany}


\vspace{0.4in}

\begin{abstract}
String theoretical arguments led to the hypothesis that the ratio of
viscosity to entropy of any physical system has a lower bound. Strongly
coupled systems usually have a small viscosity compared to weakly coupled
plasmas in which the viscosity is proportional to the mean free path. 
In the case of a one-component plasma the viscosity as a function of 
the coupling strength shows a minimum. Here we show that the ratio of 
viscosity to entropy of a strongly coupled one-component plasma is always 
above the lower bound predicted by string theory. 
\end{abstract}

\pacs{52.27.Gr}

\maketitle

\vspace{0.2in}

Recently, string theoretical (AdS/CFT) arguments indicated that
there is a lower limit for the ratio of shear viscosity $\eta$
to the entropy density $s=S/V$ \cite{kovtun},
\begin{equation}
\frac{\eta}{s}\geq \frac{\hbar}{4\pi k} = 6.08 \times 10^{-13}\> {\rm K\, s}, 
\label{e1}
\end{equation} 
where $k$ is the Boltzmann constant. This lower bound should hold for all 
relativistic quantum field theories such as QED, and
therefore should be relevant also for non-relativistic systems \cite{kovtun}. 
This prediction is in particular of interest for strongly coupled systems,
which usually have a small viscosity. A famous, widely discussed example is 
the quark-gluon plasma (QGP), which might be produced in relativistic
heavy-ion collisions \cite{kovtun,horowitz}. Other interesting examples are the
strongly coupled one-component plasma (OCP) and Yukawa systems, which
are non-relativistic, classical many-body systems. Applications for these
systems are realized in the ion component in white dwarfs and in complex 
(or dusty) plasmas.  

Here we want to estimate $\eta /s$ for the OCP as a test
for the above hypothesis (\ref{e1}). 
The OCP consists of equally charged particles 
with Coulomb interaction in a neutralizing background. The most important 
quantity describing the OCP is the Coulomb coupling parameter 
(see e.g. \cite{ichimaru})
\begin{equation}
\Gamma = \frac{Q^2}{4\pi \epsilon_0\> d\> kT},
\label{e2}
\end{equation} 
where $Q$ is the charge of the 
plasma particles, $d$ the interparticle distance, 
and $T$ the plasma temperature. If $\Gamma > 1$, one 
speaks of a strongly coupled plasma, which shows a liquid-like behavior. 
For $\Gamma > 172$, numerical simulations showed that a crystalline (solid)
phase should appear \cite{ichimaru}, which was confirmed in the case
of complex plasmas \cite{thomas}. 

Let us first consider the viscosity of the OCP.
Numerical simulations based on molecular dynamics for 
the OCP and Yukawa systems predicted a  
minimum of the shear viscosity at $\Gamma \sim 20$ \cite{saigo}.
The origin of this  minimum comes from the fact, that on one side 
the viscosity is large in the weak coupling limit because of the
large mean free path to which the viscosity is proportional 
according to the kinetic theory \cite{reif}. On the other side,
for strong coupling there is an exponential increase of the
viscosity due to caging of the particles by their neighbors
(Arrhenius law) \cite{ivlev}.

In Ref.\cite{saigo} the normalized viscosity, defined by 
\begin{equation}
\eta^*=\frac{\eta}{m\, n\, \omega_p\, d^2}
\label{e3}
\end{equation} 
was calculated. Here $m$ is the mass of the plasma particles, 
$n=N/V=3/(4\pi d^3)$ the particle number density,
and $\omega_p=(Q^2n/\epsilon_0 m)^{1/2}$ the plasma frequency. The denominator
of (\ref{e3}) can be related to the Coulomb coupling parameter as
\begin{equation}
m\, \omega_{p}\, d^2=\sqrt{3\Gamma}\, v_T\, d\, m,
\label{e4}
\end{equation} 
where $v_T=(kT/m)^{1/2}$ is the thermal velocity of the plasma particles.

A simple fit, valid in the vicinity of the minimum, of the numerical result
was given as \cite{saigo}
\begin{equation}
\eta^* = \frac{\alpha}{\Gamma}+\beta \Gamma +\gamma.
\label{e5}
\end{equation} 
where $\alpha = 0.96$, $\beta = 0.0022$, and $\gamma = -0.030$. 
The minimum of this fit is $\eta^* = 0.62$ at $\Gamma = 20.9$.

Next we consider the entropy of the OCP. Denoting 
$U$ as the total internal energy of the plasma, the 
normalized energy $u=U/(NkT)$ can be written as
$u=u_{id}+u_{ex}$, where $u_{id}=3/2$ is the ideal gas contribution
and $u_{ex}$ the so called excess energy. A very good 
fit of the Monte Carlo data for the excess energy 
in the range $1\leq \Gamma \leq 160$
was found to be \cite{slattery, ichimaru} 
\begin{equation}
u_{ex}(\Gamma)=a\Gamma+b\Gamma^{1/4}+c\Gamma^{-1/4}+d
\label{e6}
\end{equation} 
with $a=-0.89752$, $b=0.94544$, $c=0.17954$, and $d=-0.80049$.

The normalized free energy $f=F/(NkT)=f_{id}+f_{ex}$
is given by \cite{farouki}
\begin{eqnarray}
f_{ex}(\Gamma)&=&\int_1^\Gamma \frac{d\Gamma'}{\Gamma'}\> u_{ex}+f_{ex}(1)
\nonumber \\ 
&=& a(\Gamma -1)+4b(\Gamma^{1/4}-1)-4c(\Gamma^{-1/4}-1)+d\ln \Gamma+f_{ex}(1)
\label{e7}
\end{eqnarray} 
with $f_{ex}(1)=-0.4368$.

The entropy then follows from $S=(U-F)/T$. The excess contribution to the
normalized entropy, $s^*=S/(Nk)=u-f=s_{id}+s_{ex}$, reads
\begin{equation}
s_{ex}=u_{ex}-f_{ex}= -3b\Gamma^{1/4}+5c\Gamma^{-1/4}-d\ln \Gamma +d+a+4b-4c
-f_{ex}(1)
\label{e8}
\end{equation} 
The entropy of an ideal gas of indistinguishable particles \cite{diu,hansen}
is given by
\begin{equation}
s_{id}=\frac{5}{2}-\ln n-3\> \ln \Lambda = \frac{5}{2} +\ln \frac{4\pi}{3}+
3\> \ln \frac{d}{\Lambda},
\label{e9}
\end{equation} 
where $\Lambda = (2\pi \hbar^2/mkT)^{1/2}$ is the de Broglie wave length.
It should be noted that $d$, $\Lambda$, and $\Gamma $ are not independent 
for a fixed mass and charge of the plasma particles.

Now we find for the ratio of viscosity to entropy density
\begin{equation}
k \frac{\eta}{s}=R(\Gamma)\> v_T\> m\> d,
\label{e10}
\end{equation} 
where
\begin{equation}
R(\Gamma)=\frac{\sqrt{3\Gamma}\eta^*}{s^*}.                                     
\label{e11}
\end{equation} 
$R(\Gamma)$ depends only on $\Gamma$ apart from the weak
logarithmic dependence on $d/\Lambda$ in $s_{id}$. Hence, for
having a small ratio of $\eta/s$ one should look for the
minimum of $R(\Gamma)$ and a small value for $v_T\> m\> d$ at the same
time. The latter, however, is bounded from below by the assumption
of a classical approximation, which means that $\Lambda \leq d$ leading to
$v_T\> m\> d \geq \sqrt{2\pi}\> \hbar$. Using the lower limit 
corresponding to $\Lambda =d$ for
$v_T\> m\> d$ in (\ref{e10}), 
the string theory limit (\ref{e1}) can be expressed now as
$(32\pi^3)^{1/2} R(\Gamma) >1$. Combining (\ref{e5}), (\ref{e8}), and
(\ref{e9}) the minimum of lhs is 4.89 at $\Gamma =12$, i.e. the
prediction (\ref{e1}) is fulfilled. As a matter of fact, a OCP at the
classical limit ($\Lambda =d$) 
comes quite close to the theoretical limit (\ref{e1})
compared to other systems discussed so far \cite{kovtun, csernai}, e.g. liquid helium 
which is a factor 9 above this limit \cite{kovtun}, in particular,
if one takes into account that there is only one component. The presence of
many components in a strongly coupled plasma would lead to an increase of the 
entropy and might therefore reduce $\eta /s$ below the limit (\ref{e1})
similarily as discussed in the case of a meson gas \cite{cohen}.  

As a possible application for the OCP we will discuss now
ions in a white dwarf. Since the electrons are highly degenerated,
the electron-ion interaction can be neglected and the 
electrons are only a neutralizing background. Therefore the OCP is 
good approximation in this case \cite{vanhorn}.
We assume the following typical properties of a white dwarf \cite{camenzind}:
In the interior it is composed of carbon and oxygen, i.e. we
use $Z=7$ and $m=14$ amu. The mass density is $\rho = n m \simeq 10^9$ 
kg/m$^3$, from which $n=4.3\times 10^{34}$ m$^{-3}$ and $d=1.8 
\times 10^{-12}$ m follows. The temperature lies in the range
$T=10^6 - 10^8$ K depending on the age of the white dwarf cooling down after
its formation. Hence the Coulomb coupling parameter of the ion component 
in a white dwarf is in the range $\Gamma \simeq 5 - 500$. Note that
for large values of $\Gamma$ the ions can be in crystalline state 
leading to speculations about a diamond core \cite{camenzind}.
Using as an example $T=10^7$ K corresponding to $\Gamma =45$ , it follows that
$\Lambda = 1.5 \times 10^{-13}\> {\rm m} \ll d$ (classical regime) and
$v_T=7.7 \times 10^4$ m/s. Hence
\begin{equation}
\frac{\eta}{s} = 2.6 \times 10^{-10}\> {\rm K\, s}, 
\label{e12}
\end{equation} 
which is about a factor of 400 above the string theory limit (\ref{e1}), which is
similar to water under normal conditions \cite{kovtun}.

Considering Yukawa systems would be a possible extension of this investigation.
For the equation of state (total energy, free energy, entropy) in this case 
see Ref.\cite{farouki} and Ref.\cite{saigo} for the viscosity. In the case
of a Yukawa interaction a new parameter, $\kappa=d/\lambda_D$, appears, where
$\lambda_D$ is the Debye screening length. Now the constants $\alpha$ etc. in
(\ref{e5}) and $a$ etc. in (\ref{e8}) depend on $\kappa$. An important
application are complex plasmas. Note that in this case the particles
are distinguishable, i.e. the entropy density is non-extensive because
$s_{id}=\frac{3}{2}+\ln V-3\> \ln \Lambda$ \cite{diu}. Also there might be 
other contributions to the entropy coming for example from charge fluctuations
\cite{pandey}. Therefore it would be of interest to use complex plasmas 
as a test for the string theory prediction (\ref{e1}).

Summarizing, the strongly coupled OCP obeys the string theoretical 
prediction for an lower limit of the ratio of viscosity to entropy, although 
it comes closer to it then most other systems considered so far. 
Multi-component and complex plasmas might challenge the string theory 
hypothesis.

\bigskip


\end{document}